\newcommand\sps{\space\space\space\space}
 \def\selectedoptions{final}

\documentclass[
   \selectedoptions
  ]
  {aipproc}

\def\selectedlayoutstyle{6x9}
\layoutstyle\selectedlayoutstyle

\SetInternalRegister\hbadness{8000} 

\newcommand\doingARLO[2][]{%
  \ifx\mmref\undefined #1\else #2\fi
}

\begin{document}

\title 
{Some Aspects of Fundamental Symmetries and Interactions}

\classification{}
\keywords{antiprotons, fundamental symmetries, fundamental interactions, precision measurements }

\author{Klaus P. Jungmann}{
  address={Kernfysisch Versneller Instituut, Rijksuniversiteit Groningen, Zernikelaan 25, 
                   9747 AA Groningen, The Netherlands},
  email={jungmann@kvi.nl},
  thanks={}
}

\copyrightyear  {2005}

\begin{abstract}
The known fundamental symmetries and interactions are well described by the Standard Model.
Features of this powerful theory, which are described but not deeper explained, are addressed
in a variety of speculative models. Experimental tests of the predictions in such approaches can be either 
through direct observations at the highest possible accelerator energies or through precision 
measurements in which small deviations from calculated values  within the Standard Model are searched 
for. Antiproton physics renders a number of possibilities to search for new physics. 
\end{abstract}

\date{\today}

\maketitle

In modern physics symmetries play an important and central role. 
Global symmetries are connected with conservation laws and local symmetries 
give rise to  forces \cite{Lee_56}. Today four fundamental interactions are
known:
Electromagnetism, 
Weak Interactions,
Strong Interactions, and 
Gravitation.
They are considered fundamental, because all
observed dynamic processes in nature can be traced back to one or
a combination of them.

Electromagnetic, Weak and Strong 
Interactions can be described in one single theory to astounding precision,
the Standard Model (SM) \cite{Marciano_2004}. However, the SM leaves yet several
intriguing questions unanswered. Among those are, e.g., 
the number of fundamental particle generations,
the hierarchy of the  fundamental fermion masses,
and the  physical origin of the observed breaking of discrete 
symmetries in weak interactions,
e.g. of parity (P), of time reversal (T) and of 
combined charge conjugation and parity (CP),
although the experimental findings can be well described.
Further, the large number of some 27 free parameters 
in the SM \cite{Marciano_2004} is unsatisfactory. 

In order to explain some of the not well understood features in the SM,
searches for 
yet 
unknown interactions  are very important. Such forces are suggested by a variety of speculative 
models in which extensions to the present standard theory are introduced.
There are models with left-right symmetry, 
fundamental fermion compositeness, new particles, leptoquarks, 
supersymmetry, technicolor and many more.
It is  also a major goal to find a unified 
quantum field theory which includes all the four known
fundamental forces. For this, a satisfactory quantum 
description of gravity remains yet to 
be found. In this lively area of actual activity  string or membrane theories 
provide  promising approaches. 

Without secure experimental verification in the future any of these speculative
theories will remain without status in physics, independent of
the mathematical elegance and their appeal. Experimental searches
for predicted unique features of those models are therefore essential
to steer theory towards a better and deeper understanding of the 
fundamental laws in nature. Two main lines of experimental approach are followed at present:
(i) the direct observation of new particles and processes at the highest energies achievable and
(ii) the precise measurement of quantities which can be accurately calculated within the SM
and where 
a discrepancy between theory and experiment would indicate
new physics. Both methods deliver complementary information.

In this paper we will discuss some recent developments in the field to frame
numerous present and future activities in antiproton ($\overline{p}$) research  \cite{Jungmann_2005}.

\section{Fundamental Fermion Properties}

\subsection{Neutrinos - Mixing, Masses and their Nature}
The reported evidence for neutrino ($\nu$) oscillations 
\cite{neutrino_expts} strongly indicate finite $\nu$ masses
\cite{neutrino_reviews}. In particular, the neutrino mass eigenstates  ($\nu_1$, $\nu_2$, $\nu_3$) 
are mixed in the observed  flavor states ($\nu_e$, $\nu_{\mu}$, $\nu_{\tau}$).  
Among the recent discoveries are the surprisingly large
mixing angles $\Theta_{12}$ and $\Theta_{23}$.
The mixing angle $\Theta_{13}$, the phases for possible CP-violations, and 
the question whether $\nu$'s are Dirac or Majorana particles
rank among the top issues in neutrino physics.
Since the oscillation experiments only yield differences of squared masses,
a direct measurement of a $\nu$ mass is highly desirable\cite{neutrino_reviews}. 
To address the mixing angle and the CP-violation new neutrino beam experiments
are proposed using a neutrino factory or $\beta$-beams and large new Cherekov detectors \cite{CERN_PD_2004}.
The new accelerators needed might well be suitable
to produce intense beams of antiprotons ($\overline{p}$).

The best neutrino mass limits have been extracted from measurements of the
tritium $\beta$-decay spectrum close to its endpoint.
Spectrometers based on Magnetic  Adiabatic Collimation combined with an Electrostatic filter
(MAC-E technique) and found $m(\nu_e) < 2.2~eV$ \cite{MainzTroitzk}. 
A new experiment exploiting the same technique, KATRIN \cite{KATRIN_2005}, is presently prepared
in Karlsruhe, Germany. It aims for about one order of magnitude improvement.
This yields sensitivity to the mass range
where a  finite effective neutrino mass value of between 0.1 and 0.9 eV was
claimed from a signal in neutrinoless double $\beta$-decay in $^{76}$Ge 
\cite{Klapdor_2004} with a 4.2 $\sigma$ effect. 
{Neutrinoless double $\beta$-decay is only possible
for Majorana neutrinos. This decay gives  not only the best known key to the question
of the neutrino nature, it also is the only approach at present towards finding total
lepton number violation. It is addressed in several different experiments.}

\subsection{Quarks - Unitarity of Cabbibo-Kobayashi-Maskawa-Matrix}

The mass and weak eigenstates of the u, s and b quarks  are different and related to each other
by a $3 \times 3$ unitary matrix, the Cabbibo-Kobayashi-Maskawa (CKM) matrix \cite{PDG}. Non-unitarity
would be an indication of physics beyond the SM and could be
caused by many possibilities, including the existence of more than
three quark generations. The best test of unitarity results from the first row of the 
CKM matrix through
${\rm V}_{ud}|^2 + |{\rm V}_{us}|^2 + |{\rm V}_{ub}|^2 = 1 -\Delta $,
where the SM predicts $\Delta$ to be zero. With 
the present uncertainties only the elements V$_{ud}$ and V$_{us}$ play a role. 
V$_{ud}$ can be extracted most accurately from  ft values of 
superallowed $\beta$-decays, neutron
decay and pion $\beta$-decay, which all are presently measured.
V$_{us}$ can be extracted from K decays and in principle also from
hyperon decays.
Some 2.5 $\sigma$ deviation from unitarity 
had been persistently reported \cite{PDG, Abele_2004}. 
Recent analysis of the  subject  has revealed overlooked inconsistencies in the overall picture
 \cite{Czarnecki_2004} and at this time new determinations of
V$_{us}$ in several K-decay experiments \cite{Ellis_2004} 
together with V$_{ud}$ from nuclear $\beta$-decay \cite{Hardy_2005}
confirm  $\Delta=0$ and the unitarity of the CKM matrix.

\section{Discrete Symmetries}
Violations of the discrete symmetries P, C and T as well as for CP have been directly observed 
 \cite{PDG} in weak interactions. The  combined CPT symmetry is not known
to be violated \cite{CPT_0}. Assuming CPT being conserved CP violation implies 
T-violation.

\subsection{Parity}
The observation of neutral currents together with the
measurements of parity non-conserva\-tion in atoms were
important to establish the validity of the SM.  
Processes over 10 orders in momentum transfer - from atoms to highest energy scattering -
are described by the same electro-weak parameters. This 
is
one of the biggest successes in physics.

At the level of highest precision electro-weak experiments \cite{Hughes_2004}
questions arose recently, which ultimately may call for a refinement of theory.     
The predicted running of the weak mixing angle $sin^2 \Theta_W$
appears  not to be in agreement with observations \cite{Marciano_2004,Czarnecki_1998}.
If the value of   $sin^2 \Theta_W$  is fixed at the Z$^0$-pole, deep inelastic neutrino scattering
at several GeV appears to yield a considerably higher value. 
A new round of experiments is being started with the Q$_{weak}$ experiment \cite{Qweak} 
at the  Jefferson Laboratory in the USA.
In the same context  a reported disagreement of atomic parity violation in Cs  \cite{Wieman_1999} 
has disappeared after a revision of atomic theory.
  For atomic parity violation \cite{Bouchiat_1999} 
in principle higher experimental accuracy will be possible from experiments using
Fr isotopes \cite{Atutov_2003,Gomez_2004} or single Ba or Ra ions in radiofrequency traps \cite{Fortson}.

At the CERN LEAR facility
$\overline{p}$ x-rays from atoms in which a $\overline{p}$ was 
captured were utilized to obtain information on the neutron mean square 
nuclear radii \cite{Trzcinska_2004}. The achieved accuracy
is presently limited by nuclear theory. Neutron distributions
are expected to be the limiting factor in the theory for the upcoming round of precision
experiments on atomic parity violation, in particular for some 
Fr and Ra isotopes. At a combined radioactive beam and
antiproton facility one can expect experiments to determine the neutron radii with
sufficiently high accuracy for the theory of atomic parity violation.

\subsection{Time Reversal and CP Violation}
Searches for new sources of  CP-violation 
are particularly motivated because of a  possible relation 
to the  matter-antimatter  asymmetry in the universe.   
Sakharov \cite{Sakharov_1967} suggested that the 
observed dominance of matter could be explained via CP-violation in the early universe 
at thermal non-equilibrium and 
baryon number violating
processes.
(In a later model of Bertolami et al. \cite{Bertolami_1996} only CPT violation 
and baryon number violation are needed.)
 CP violation as described in the SM 
is insufficient to satisfy the needs of this model.
Permanent Electric Dipole Moments (EDMs) and  certain
correlation observables  in $\beta$-decays offer opportunities 
to find new sources of CP-violation.

An EDM of any fundamental particle
violates both P and T  invariance. 
EDMs  for all particles are caused by CP violation as it is known from 
the K systems through higher order loops. These are at least 4 orders of magnitude below the
present experimentally established limits. A number of speculative models
foresees EDMs as large as 
the present experimental limits just allow. 
EDMs were
searched in various systems with different sensitivities 
(for details see e.g. \cite{Jungmann_2005a}). In composed systems such as molecules
or atoms fundamental particle EDMs of constituents may be
significantly enhanced \cite{Sandars_2001}. Particularly in polarizable
systems  large internal fields can exist and can be exploited.

There is no preferred system to search for an EDM
\cite{Jungmann_2005a}. In fact,
many objects need to be examined, because depending
on the underlying process different systems have
in general quite significantly different susceptibility
to acquire an EDM through a particular mechanism.
An EDM may be found an 'intrinsic property' of
an elementary particle as we know them, because the underlying 
mechanism is not accessible. However, it can also
arise from CP-odd forces between the constituents,
e.g., between nucleons in nuclei or between nuclei and
electrons. Such EDMs could be much larger than those
expected for elementary particles originating within the usual
popular SM extensions. 

There are recent novel developments: 
(i) For Ra isotopes  a unique $e^-$ EDM enhancement 
was predicted in certain atomic states \cite{Dzuba_2001,Jungmann_2002}
and   in certain nuclei (dynamic) octupole deformation may enhance the effect of a nucleon EDM
substantially \cite{Engel_2004}. 
(ii) A very novel idea was introduced  for searching an 
EDM of a charged particle. The  motional electric field is exploited, which charged particles at relativistic speeds 
experience in a magnetic storage ring. This method can be applied for muons \cite{Farley_2004} 
to obtain information on the second  generation of particles without strangeness \cite{Feng_2004,Babu_2000} 
or  deuterons \cite{Semertzidis_2004}to yield higher sensitivity to quark chromo EDMs than in neutrons\cite{Liu_2004}.
This new method may allow also to search sensitively for a $\overline{p}$ EDM.
(iii) Molecules such as PbO became experimentally accessible with a high potential to
find an electron EDM \cite{deMille_2004}.

A different approach to T-violation comes form nuclear $\beta$-decay.
In standard theory the structure of weak
interactions is V-A, which means there are vector (V) and axial-vector (A) 
currents with opposite relative sign causing a left handed structure 
of the interaction and parity violation \cite{Herczeg_2001}.
Other possibilities like scalar, pseudo-scalar and tensor 
interactions which might be possible would be clear 
signatures of new physics.
The spectrum of present 
searches includes $\beta$-asymmetry measurements and 
measurements of $\beta$-$\nu$ correlations (see e.g. \cite{Jungmann_2005}),
where in particular T-violation could be observed.
 
It should be noted that $\overline{p}$ physics has made significant contributions
towards understanding CP-violation and the direct observation of T-violation,
e.g., in the context of the CP-LEAR experimental programme \cite{CPLEAR}.

\subsection{CPT Invariance Tests and Properties of Known Basic Interactions }

The invariance of physical processes under a combined CPT transformation
relates to a number of basic physical phenomena, such as Lorentz invariance and the
(non-) existence of a preferred frame of reference, the equality of particle and antiparticle
properties, spin, the existence of fermions and bosons only and many more \cite{CPT_0}.
Quite often the results of searches for CPT violation are expressed in small relative numbers,
such as, e.g., a limit on a relative deviation of particle properties \cite{PDG}. Here 
some freedom in the choice of which quantities are compared is frequently used to obtain
some small number. However, an interaction based comparison must be considered more 
physical. To this extent a theoretical model has been proposed by Kostelecky and co-workers
\cite{Kostelecky}. It allows to compare different experimental approaches 
in a single theory based on the interaction strength of a possible CPT violating term
in the Lagrangian. Such terms are treated perturbatively in systems which otherwise
 can be described well in standard theory. Therefore these measurements are 
intimately connected to the accurate determination of particle properties and fundamental constants.

\subsubsection{Trapped Charged Particles}
Trapping and storing of charged particles in combined magnetic and electric fields
has been very successfully applied for obtaining properties of 
the respective species and for determining fundamental constants.
Most accurate results were obtained from single trapped and cooled charged particles.
The comparison of electron ($e^-$) and positron ($e^+$), positive and negative muon 
($\mu^+, \mu^-$
), and proton ($p$)  and antiproton ($\overline{p}$) has already reached an impressive
level of precision.

The magnetic anomaly of fermions $a=\frac{1}{2}\cdot(g-2)$
describes the deviation of their magnetic g-factor
from the value 2 predicted in the Dirac theory. It
could be determined for single electrons and positrons in Penning traps
by Dehmelt and his coworkers to 10~ppb \cite{Dyc_90}
by measuring the cyclotron frequency and its difference to the 
spin precession frequency (g-2 measurement).
The good agreement for the magnetic anomaly for electrons and positrons is
considered the best CPT test for leptons \cite{PDG,Kostelecky}.
Accurate calculations involving
almost exclusively the "pure" Quantum Electrodynamics (QED)
of electron, positron and photon fields allow the most precise
determination of the fine structure constant $\alpha$ \cite{Kinoshita_2004,Kinoshita_1990}
by comparing experiment and theory for the electron magnetic anomaly
in which $\alpha$ appears as an expansion coefficient.
One order of magnitude improvement appears  possible \cite{Gabrielse_2005}
with a new experimental approach (also involving single cooled trapped particles) 
which aims for
reducing the effect of cavity QED, the major systematic contribution to the 
previous experiment \cite{Brown_1986}.

Muons have been stored in a series of measurements at CERN and BNL in magnetic 
storage rings with weak electrostatic focusing. These devices are conceptually
equivalent to Penning traps.
The latest g-2 experiment  \cite{Bennet_2004}
yields  values for $\mu^+$ and $\mu^-$ which agree
at the 0.7 ppm level in accordance with CPT. 
The muon is by a factor $(\frac{m_{\mu}}{m_e})^2$  more sensitive
to heavy particles compared to the electron. The muon g-2 measurements are
sensitive to new physics involving heavy particles
at 40,000 times lower experimental precision than would be needed for $e^-$. Whether
the present muon experimental results are in agreement with standard theory remains an
open question, as not sufficiently accurate values for corrections due to known
strong interaction effects exist yet.

For the first time in 1986  $\overline{p}$'s could be trapped in a cylindrical Penning trap  
 after moderation of a $\overline{p}$ beam from LEAR at CERN\cite{Gabrielse_1986}.
Effective moderation of MeV  $\overline{p}$'s  and their capture were very important, 
which has led to detailed  studies of the range differences when  $p$'s and $\overline{p}$'s 
are slowed down in matter,  known as the Barkas effect\cite{Gabrielse_1989_a}.  
Further, electron cooling is essential and could be demonstrated already in the early 
experiments \cite{Kells_1986}.
In a series of measurements in
which the cyclotron frequencies were measured the accuracy of the charge to mass ratio
for $\overline{p}$'s could be improved and compared to the proton
value. The best results were achieved when a single H$^-$ ion and 
a single $\overline{p}$ where measured alternatively in the same trap \cite{Gabrielse_1999}.
At present these experiments are interpreted as a CPT test for
$p$ and $\overline{p}$ at the level of $9\times 10^{-11}$.
A new experiment has been proposed to measure  the magnetic g-factor of the
$\overline{p}$ using single particle trapping. Similar to measurements on
single electrons and positrons the cyclotron and spin precession frequencies 
shall be determined. One expects for the comparison of $p$ and $\overline{p}$
g-factors an improvement by a factor of $10^6$ \cite{FLAIR_2005}.

\subsubsection{Hydrogen and Antihydrogen}  

Precision spectroscopy of hydrogen and its isotopes (including the exotic ones
like positronium ($e^+e^-$) and muonium ($\mu^+e^-$) atoms) has confirmed
bound state QED at a high level of precision and contributed together
with numerous precision experiments to the set of well established fundamental constants
\cite{Mohr_2005}; among those are the Rydberg constant as the best measured constant,
$\alpha$ and the $\mu$ mass and magnetic moment.

The reproducibility of high precision in atomic hydrogen laser spectroscopy
allowed to set a   limit on the time variation of \protect{$\alpha$} from
two series of repeated measurements of the 1s-2s energy
difference long time where 
$\frac {\partial \alpha}{\partial t} / \alpha = \frac{\partial}{\partial t}(\ln \alpha)
 = (-0.9\pm2.9)\times 10^{-15} y^{-1} $ 
could be established \cite{Fischer_2004}.
An analysis of these long term experiments reveals
that they were operated just at the border at which 
 systematic errors are understood in optical spectroscopy, i.e., 
 at the \protect{$10^{-13}$} level of  relative accuracy.

Antihydrogen ($\overline{{\rm H}}$) was produced first  at CERN in 1995
\cite{Baur_1996}. The atoms were fast as the production mechanism required $e^+e^-$ 
pair creation when  $\overline{p}$'s were passing near heavy nuclei. A small
fraction of the $e^+$ form a bound state with the $\overline{p}$.
The experiment was an important step forward showing that  a few $\overline{{\rm H}}$
could be produced. Unfortunately, the speed of the atoms does not allow any
meaningful spectroscopy. Later  a similar experiment was carried out at FERMILAB 
\cite{Blanford_1998}.

The successful production of slow $\overline{{\rm H}}$ was first reported by the
ATHENA collaboration in 2002 \cite{Amoretti_2002} and shortly later also by the ATRAP collaboration
\cite{Gabrielse_2002}. Both experiments use combined Penning traps in which first
$e^+$  and $\overline{p}$ are stored separately and cooled. The atoms form when both species 
are brought into contact by proper electric potential switching in the combined traps.
The detection in ATHENA  relies on diffusion of the neutral atoms out of the interaction volume and
the registration of $\pi$'s  which appear when the  atoms annihilate on contact with matter walls of the
container. In ATRAP the hydrogen atoms are re-ionized in an electric field 
and the $\overline{p}$'s are observed using 
a capture Penning trap. 
Most of the atoms are in excited states (n > 15) which can be seen from the fact that their physical size
is above 0.1 $\mu$m \cite{Gabrielse_2004}. 
For spectroscopy the atoms need to be in 
states with low $n$, 
preferentially the ground state. The production of such states is a major goal 
of the community for the immediate future.
The kinetic energy of the produced $\overline{{\rm H}}$ atoms is of
order 200 meV corresponding to a velocity of $6 \times 10^4$ m/s . 
This is a factor of 400 above the value where neutral atom traps can hold them.
Therefore cooling such atoms or identifying a production mechanism for colder 
$\overline{{\rm H}}$ are a  central topic.

For laser cooling of $\overline{{\rm H}}$  a continuous laser at the H Lyman-$\alpha$
frequency for
$\overline{{\rm H}}$ cooling  has recently been developed \cite{Eikema_2001}. 
One hopes to 
achieve  the photo-recoil limit of 1.3 mK. Recently a promising new method was 
demonstrated to obtain  $\overline{{\rm H}}$. It uses resonant 
charge exchange with excited positronium to obtain $\overline{{\rm H}}$ atoms with 
essentially the same velocities as the $\overline{p}$'s in the trap which can
be made rather low by cooling. \cite{Gabrielse_2004,Storry_2004}. 

A main motivation to perform precision spectroscopy on $\overline{{\rm H}}$
is to test  CPT invariance.  There are two electromagnetic transitions which offer a
high quality factor  and therefore promise high experimental precision
when H and $\overline{{\rm H}}$ are compared:
the 1s-2s two-photon transition at frequency $\Delta \nu_{1s-2s}$ 
and the ground state hyperfine splitting $\Delta \nu_{HFS}$, which both have within the SM
in addition to the leading order contributions from QED, nuclear structure, weak and strong interactions,
%
$\Delta \nu_{1s-2s} = \frac{3}{4} \times R_{\infty} + \varepsilon_{QED}
                                      + \varepsilon_{nucl} + \varepsilon_{weak} + \varepsilon_{strong}
                                      + \varepsilon_{CPT} $
%
and 
%
$\Delta \nu_{HFS} = const\times \alpha^2 \times R_{\infty} + \varepsilon_{QED}^*
                                      + \varepsilon_{nucl}^* + \varepsilon_{weak}^* + \varepsilon_{strong}^*
                                      + \varepsilon_{CPT}^*.$
It is assumed that only CPT violating contributions
exist from interactions beyond the SM. If one assumes that $\varepsilon_{CPT}$ and
$\varepsilon_{CPT}^*$ are of the same order of magnitude, the relative contribution is larger by order
$\alpha^{-2}\approx 2\times 10^4$  for $\Delta \nu_{HFS}$. Further one can speculate that a new 
interaction may be of short range (contact interaction), which also favors 
measurements of $\Delta \nu_{HFS}$. Such an experiment has been recently proposed.
It utilizes a cold $\overline{{\rm H}}$ atom beam and has sextupole state selection magnets in a 
Rabi type  atomic beam experiment \cite{Widmann_2004_a}.
For both experiments temperature of the atoms and statistics governs the reachable precision, i.e. the
atoms should be as cold as possible and one should use as many as possible atoms.

\subsubsection{Gravitational Force on $\overline{{\rm H}}$}
One of the completely open questions in physics concerns the sign of gravitational interaction
for antimatter. It can only be answered by experiment. A proposal \cite{Walz_2003} exists 
 in which the deflection of a horizontal cold beam is
measured in the earth's gravitational field. The experiment plans on a number of
modern state of the art atomic physics techniques like sympathetic
cooling of $\overline{{\rm H}}^+$ ions  by, e.g., Be$^+$ ions in an ion trap
to achieve the neccessary  low temperatures of some 20~$\mu$K . 
After  pulsed laser photo-dissociation of the ion into $\overline{{\rm H}}$  and a $e^+$ the
neutral atoms can then leave the trap. The atom's ballistic path can be measured.

\subsubsection{Antiprotonic Helium}

The potential of antiprotonic helium for precision measurements in the field of 
fundamental interaction research was realized shortly after it had been discovered that
$\overline{p}$'s stopped in liquid or gaseous helium exhibit long lifetimes and do not
rapidly annihilate with nucleons in the helium nucleus \cite{Yamazaki_2002}. 
This can be explained, if one assumes that the $\overline{p}$'s  are captured in metastable states
of high principal quantum number $n$  and high angular momentum $l$ , with $l\approx n$ 
 \cite{Nakamura_1994}. The capture happens  typically
at $n \approx \sqrt{M^*/m_e} \approx 38$, where $M^*$ is the reduced mass of
the ($\overline{p}$He) bound system.

With laser radiation  the $\overline{p}$'s in these atoms can be transferred into states where Auger 
de-excitation can take place. In the resulting  H-like system Stark mixing with 
s-states results in nuclear  $\overline{p}$ absorption and annihilation which is 
signaled by emitted pions. This way a number of transitions could be induced
and measured with continuously increasing accuracy over the past decade.
A precision of $6 \times 10^{-8}$ has been reached for the transition frequencies 
\cite{Hori_2001}, which has been stimulating for improving three-body QED 
calculations.  It should be noted that with the high principal quantum numbers
for the $\overline{p}$ the system shows also molecular type character
\cite{Yamazaki_2000}.

Among the spectroscopic successes the laser-microwave double resonance
measurements of hyperfine splittings of $\overline{p}$ transitions could be 
measured \cite{Widmann_2002}. There is agreement with QED theory
\cite{Korobov_2001} at the $6 \times 10^{-5}$ level which can be interpreted as
a measurement of the antiprotonic bound state g-factor to this accuracy.
Hyperfine structure measurements in antiprotonic helium
offer the possibility to measure the magnetic moment of the $\overline{p}$.

The very good agreement of the QED calculations with the
measurements of several transitions can be exploited to extract
a limit on the equality of the charge$^2$ to mass ratio for proton and $\overline{p}$.
Combined with the results of cyclotron frequency measurements \cite{Gabrielse_1999}
in  Penning traps one can conclude  that masses and charges of  proton and $\overline{p}$ 
are equal within $6\times 10^{-8}$ in full agreement with
 expectations based on the CPT theorem \cite{Hori_2003}.  The collaboration 
estimates that a test down to the 10 ppb level should be possible.

\section{Present Contributions of Antiproton Physics to Fundamental Symmetry and Interaction Research }
The antiproton research programmes have made already a number of important contributions
to test fundamental symmetries and to verify precise calculations. With cyclotron frequency 
measurements of a single trapped $\overline{p}$ and  with precision spectroscopy of antiprotonic helium ions
stringent  CPT tests could be performed on $\overline{p}$ parameters. With precise measurements
of $\overline{p}$ in antiprotonic helium atomcules the bound state QED
three-body systems could be challenged, which has led to significant advances in theory already.
 With antiprotonic heavy atoms
new input could be provided to obtain neutron radii of nuclei, a method that may become important 
for the theory of atomic parity violation. 
The differences in proton/ antiproton interactions with matter could expand on similar work
with other particle/ antiparticle systems.

 $\overline{{\rm H}}$ atoms have been produced by two independent collaborations. 
Precision spectroscopy of these atoms will depend on the availability of atoms in the ground state, 
the successful  cooling of the systems to below the 100 ~$\mu$eV range and their confinement in neutral particle traps. 
Work is in progress towards tests of the CPT invariance using the 1s-2s interval or the ground 
state hyperfine splitting. 
 A comparison with other exotic atom experiments 
shows that one must allow for sufficient time to develop the neccessary understanding
of production mechanisms and one must allow time 
for improving the techniques. The experiments will benefit in their speed of progress and in their 
ultimate precision
from future slow $\overline{p}$ sources of significantly improved particle fluxes and brightness 
as compared to
today's only operational facility. 
Future possible D$^0$ decay experiments (e.g. in the context of PANDA at  FAIR) have a potential to
discover new sources of CP-violtion or violation of charged  lepton family number.
In particular for tests of antimatter gravity cold  $\overline{{\rm H}}$ atoms have a unique
potential for a major discovery.

The ongoing and planned experiments bear a robust discovery potential for new physics,
in particular when searching for CPT violation. We can look forward to future precision 
$\overline{p}$ experiments continuing  to deepen insights in fundamental interactions 
and symmetries, providing  important data and parameters within standard theory  and 
providing  improved searches for new physics in particular with the availability of 
better  $\overline{p}$ sources at CERN \cite{Yamazaki_2005,ELENA_2004} 
or possibly at the future F(L)AIR facility \cite{FLAIR_2005}.

\begin{theacknowledgments}
This work has been supported by the Dutch Stichting voor Fundamenteel Onderzoek der Materie (FOM)
in the framework of the TRI$\mu$P programme.
The author wishes to thank the organizers of the LEAP05 conference for providing a stimulating atmosphere and
for supporting his participation.
\end{theacknowledgments}


\begin{thebibliography}{9999}


\bibitem{Lee_56}           	T.D. Lee and  C.N. Yang, Phys. Rev. {\bf 98}, 1501  (1955)
\bibitem{Marciano_2004}    	W.J. Marciano, hep-ph/0411179 (2004)
\bibitem{Jungmann_2005}     K. Jungmann, Nucl.Phys.{\bf A751}, 87, (2005), nucl-ex/0501029 and
                                                    proceedings of the worksop ''Physics with Ultra Slow Antiproton Beams'',
                                                    Riken, Japan (2005) 
\bibitem{neutrino_expts}       A.B. McDonald, Nucl.Phys.{\bf A751},53 (2005) and
                                                    K. Nakamura, Nucl.Phys.{\bf A751}, 67 (2005)
\bibitem{neutrino_reviews} 	for a review see, e.g.:Y. Grossmann,hep-ph/0305245 (2003)
\bibitem{CERN_PD_2004}   A. Blondel et al., ''Physics with a Multi-MW Proton Source'', 
                                                   CERN-SPSC-2004-024 (2004);								
\bibitem{MainzTroitzk}       V.M. Lobashov et al., Phys. Lett. {\bf B460}, 227 (1999);
                                                  Ch. Kraus et al. Eur.Phys.J.{\bf 40}, 447 (2005)  
\bibitem{KATRIN_2005}      J. Angrik et al., FZ Karlsruhe, FZKA 7090 (2005)
\bibitem{Klapdor_2004}       H.V. Klapdor-Kleingrothaus, Phys. Lett. {\bf B586}, 198 (2004)
\bibitem{PDG}                      S. Eidelmann et al., Phys. Lett. {\bf B 592}, 1 (2004);
			            K. Hagiwara et al., Phys. Rev. {\bf D 66},  010001 (2002) 
\bibitem{Abele_2004}           H. Abele et al., Eur. Phys. J. {\bf C 33}, 1 (2004) 
\bibitem{Czarnecki_2004}   A. Czarnecki, W. Marciano and A. Sirlin,  
                                                 Phys.Rev.{\bf D70}, 093006 (2004)and references therein 
\bibitem{Ellis_2004}            J. Ellis, hep/ph-0409360 (2004)
\bibitem{Hardy_2005}         J.C. Hardy, Nucl.Phys. {\bf A752}, 101 (2005);
                                                 J.C. Hardy and I.S. Towner, nucl-th/0412056 (2004)
\bibitem{CPT_0} J. Schwinger, Phys. Rev. {\bf 82}, 914 (1951);
	   G. Lueders, Dansk. Mat. Fys. Medd. {\bf 28}, 17 (1954),Ann.Phys. {\bf 2}, 1 (1957);
               W. Pauli, in: {\it Niels Bohr and the Development of Physics} , McGraw-Hill (1955),
                Nuovo Cimento {\bf 6}, 204 (1957)
\bibitem{Hughes_2004}      E.W. Hughes, in: ''In Memory of Vernon Willard Hughes'',
                           E.W. Hughes and F. Iachello (eds.), World Scientific, Singapore,
                           p. 154 (2004)   
\bibitem{Czarnecki_1998}   A. Czarnecki and W. Marciano, Int.J.Mod.Phys. {\bf A13}, 
                                                  2235 (1998)  and references therein
\bibitem{Qweak}            D. Armstrong et al, proposal E02-020 to Jefferson Lab (2002)
\bibitem{Wieman_1999} S.C. Bennett and C.E. Wieman, Phys.Rev.Lett.{\bf 82}, 2484 (1999
\bibitem{Bouchiat_1999}    ''Parity Violation in Atoms and Polarized Electron Scattering'', 
                            B.F. Bouchiat and M.A. Bouchiat (eds.), Worls Scientific, 
                           Singapore (1999); see also: J. Guena, M. Lintz and M.A. Bouchiat, 
		Mod.Phys.Lett. {\bf A}, in print (2005)
\bibitem{Atutov_2003}      S.N. Atutov et al., Hyperfine Interactions {\bf 146-147}, 83 (2003)
\bibitem{Gomez_2004}       E. Gomez et al., physics/0412073 (2004)
\bibitem{Fortson}          N. Fortson, in: loc. cit. \cite{Bouchiat_1999} p. 244 (1999) 
\bibitem{Trzcinska_2004} A. Trzcinska et al., Nucl. Instr.  Meth. {\bf B214}, 157 (2004)
\bibitem{Sakharov_1967}    	A. Sakharov,JETP {\bf 5}, 32 (1967); M. Trodden, 
                           		Rev. Mod. Phys. {\bf 71}, 1463 (1999) 
\bibitem{Bertolami_1996} 	O. Bertolami et al., Phys.Lett.{\bf B395},178 (1997) 
\bibitem{Jungmann_2005a}    K. Jungmann, physics/0501154 (2005) 
\bibitem{Sandars_2001}     P.G.H. Sandars, Contemp. Phys. {\bf 42}, 97 (2001)
\bibitem{Dzuba_2001}       V. Dzuba et al., Phys. Rev. A {\bf 63}, 062101 (2001)
\bibitem{Jungmann_2002}   G.P. A. Berg et al. Nucl.Instr.Meas.B 204, 532 (2003),
                                                 K. Jungmann,  Acta Phys.Polon.{\bf 33}, 2049 (2002)
\bibitem{Engel_2004}         J. Engel et al.,Phys. Rev. {\bf C 68},
                                               025501 (2003)
\bibitem{Farley_2004}      F.J.M. Farley et al., Phys. Rev. Lett. {\bf 93}, 052001 (2004)
\bibitem{Feng_2004}        W.-F. Chang and J.N. Ng, hep-ex/0307006 (2004)
\bibitem{Babu_2000}        K.S. Babu, B. Dutta and R. Mohapatra, Phys.Rev.Lett. {\bf 85}
                                             5064 (2000);
                                              B. Dutta and R. Mohapatra, Phys. Rev. {\bf D68} (2003) 113008
\bibitem{Semertzidis_2004} Y. Semertzidis et al., AIP Conf. Proc. {\bf 698}, 200 (2004)
\bibitem{Liu_2004}         C.P Liu and R.G.E. Timmermans, nucl-th/0408060 (2004)
\bibitem{deMille_2004} D. Kawall et al., AIP Conf.Proc.{\bf 698}, 192 (2004)
\bibitem{Herczeg_2001}     P. Herczeg, Prog. Part. and Nucl. Phys. {\bf 46}, 413 (2001)
\bibitem{CPLEAR}            A.Angelopoulos et al., Phys.Rep. {\bf 374}, 165 (2003)
\bibitem{Kostelecky}    N. Russell, this conference
\bibitem{Dyc_90} R.~Van~Dyck,~Jr.,in {\it Quantum Electrodynamics}, T.~Kinoshita,
                 ed.,  p. 322, World Scientific (1990)
\bibitem{Kinoshita_2004} T. Kinoshita and M. Nio, Phys. Rev. {\bf D70} 113001 (2004)
\bibitem{Kinoshita_1990}T. Kinoshita (ed.),  {\it Quantum Electrodynamics}, World Scientific (1990) 
\bibitem{Gabrielse_2005} G. Gabrielse, priv. comm. (2005)
\bibitem{Brown_1986} L.S. Brown and G. Gabrielse, Rev. Mod. Phys. {\bf 58}, 233 (1986)
\bibitem{Bennet_2004} G.W. Bennett et al.,  Phys. Rev. Lett. {\bf 92}, 161802 (2004)
\bibitem{Gabrielse_1986} G. Gabrielse et al., Phys. Rev. Lett. {\bf 57}, 2504 (1986)
\bibitem{Gabrielse_1989_a} G. Gabrielse et al., Phys. Rev. {\bf A 40}, 481 (1989)
\bibitem{Kells_1986} G. Gabrielse et al., Phys. Rev. Lett.{\bf 63},1360 (1989)
\bibitem{Gabrielse_1999} G. Gabrielse et al., Phys. Rev. Lett. {\bf 82}, 3198 (1999); 
\bibitem{FLAIR_2005}E. Widmann et al., Letter of Intent to GSI/FAIR (2005)
                http://www.oeaw.ac.at/smi/flair/ LOI/FLAIR-LOI-resub.pdf
\bibitem{Mohr_2005}  Peter J. Mohr, Barry N. Taylor, Rev. Mod. Phys. {\bf  77}, 1 (2005)
\bibitem{Fischer_2004}M. Fischer et al., Lecture Notes in Physics {\bf 648}, 209 (2004);
                see also: M. Fischer et al., Phys. Rev. Lett. {\bf 92}, 230802-1 (2004)
\bibitem{Baur_1996} G. Baur et al., Phys. Lett. {\bf B 368}, 251 (1996) 
\bibitem{Blanford_1998} G. Blanford et al., Phys. Rev. Lett. {\bf 80}, 3037 (1998) 
\bibitem{Amoretti_2002} M. Amoretti et al., Nature {\bf 419}, 456 (2002)
\bibitem{Gabrielse_2002} G. Gabrielse et al., Phys. Rev. Lett. {\bf 89}, 213401 (2002)
\bibitem{Gabrielse_2004} G. Gabrielse, Adv. At. Mol. Opt. Phys. {\bf 50} (2004);
                see also: P. Oxley et al, Phys, Lett. {\bf B 395}, 60 (2004); M. Amoretti et al., 
                Physics of Plasmas {\bf 10}, 3056  (2003); M. Amoretti et al.,  Phys. Rev. Lett. 
                {\bf 91}, 0055001-1 (2003); M. Amoretti et al., Phys. LEtt. {\bf B 590}, 133 (2004)
\bibitem{Eikema_2001} K.S.E. Eikema et al, Phys. Rev. Lett. {\bf 86}, 5679 (2001)       
\bibitem{Storry_2004} C.H. Storry et al., Phys. Rev. Lett. {\bf 93}, 263401(2004)
\bibitem{Widmann_2004_a} E. Wildman et al., Nucl. Instrum. Meth.{\bf B214}, 89 (2004)
\bibitem{Walz_2003} J. Walz and T. H\"ansch, Gen. Rel. Grav.{\bf 36}, 561 ( 2004)
\bibitem{Yamazaki_2002} T. Yamazaki et al., Physics Reports {\bf 366}, 183 (2002)
\bibitem{Nakamura_1994} S.N. Nakamura et al., Phys. Rev. {\bf A 49}, 4457 (1994)
\bibitem{Hori_2001} M. Hori et al., Phys. Rev. Lett. {\bf 87}, 093401 (2001)
\bibitem{Yamazaki_2000} T. Yamazaki, in: {\it The Hydrogen Atom}, 
                   S.G. Karshenboim et al. (eds.),                Springer, p. 246  (2001)
\bibitem{Widmann_2002} E. Widmann et al., Phys. Rev. Lett. {\bf 89}, 243402 (2002)
\bibitem{Korobov_2001} V. Korobov and D. Baklanov, J. Phys. B: At. Mol. Opt. Phys. {\bf 34},
                L519 (2001) and Phys. Rev. {\bf A 57}, 1662 (1998)
\bibitem{Hori_2003} M. Hori et al., Phys. Rev. Lett. {\bf 91}, 123401 (2003) 
\bibitem{Yamazaki_2005} N. Kuroda et al., Phys. Rev. Lett. {\bf 94}, 023401 (2005)

\bibitem{ELENA_2004} P. Belochitskii et al., Letter of Intent to CERN, SPSC-I-231,
                   2005-010 (2005)

\end{thebibliography}
\end{document}